\documentclass[11pt, a4paper]{article}
\usepackage{jheppub}

\begin{document}

\title{Impurity effect in a holographic superconductor}

\author[a]{Takaaki Ishii,}
\author[b]{and Sang-Jin Sin}
\affiliation[a]{Department of Physics and Astronomy, Seoul National University, Seoul 151-747, Korea}
\affiliation[b]{Department of Physics, Hanyang University, Seoul 133-791, Korea}
\emailAdd{ishiitk@yukawa.kyoto-u.ac.jp}
\emailAdd{sjsin@hanyang.ac.kr}

\abstract{We consider a holographic superconductor with homogeneous impurities added. We start with the holographic Abelian-Higgs model for s-wave superconductivity, and turn on a coupling between the gauge field and a new massive gauge field that is introduced for impurities, whose effect is examined in the probe limit. We find that the condensation of the massive gauge field is induced in the superconducting phase. When the coupling is sufficiently large, the mass gap in the optical conductivity disappears. A resonance peak is found in the conductivity for the massive vector field.}

\dedicated{{\rm SNUTP12-005}}

\maketitle

\section{Introduction}
\label{sec:Introduction}

The AdS/CFT duality is a brilliant concept that allows us to describe strongly coupled conformal field theories with their gravity duals \cite{Maldacena:1997re,Gubser:1998bc,Witten:1998qj}. Certainties of this correspondence have been confirmed in so many examples, and it has been applied to various theories with much success. It is tempting  to use this powerful tool to study strongly coupled systems  with  physical importance. 

Recently, applications to condensed matter physics have been intensively investigated.\footnote{For reviews, see \cite{Sachdev:2008ba,Hartnoll:2009sz,Herzog:2009xv,McGreevy:2009xe,Horowitz:2010gk,Sachdev:2010ch,Hartnoll:2011fn}.} For example, superconductivity is modeled in gravity duals. First, it was  pointed out that the black holes in AdS can accommodate scalar hairs \cite{Gubser:2008px}: a scalar field can have nonzero condensate outside the horizon of the black holes if temperature is sufficiently low. Based on this idea, holographic model of superconductor was then proposed in \cite{Hartnoll:2008vx}, and many generalizations were considered. (For instance, see \cite{Gubser:2008wv,Horowitz:2008bn,Hartnoll:2008kx,Horowitz:2009ij}. See also references in \cite{Sachdev:2008ba,Hartnoll:2009sz,Herzog:2009xv,McGreevy:2009xe,Horowitz:2010gk,Sachdev:2010ch,Hartnoll:2011fn}.) Since gravity duals describe strongly coupled systems, it is hoped that holographic superconductors would unveil the mechanism of high-$T_C$ superconductors.

In condensed matter systems, considering the effect of impurities is often important since their presence can drastically change the physical properties of the systems. They can be used to control the conductivity of semiconductors. In superconductors, systems can show gapless superconductivity in the presence of impurities. Apart from such practical applications, it also  generates physically interesting phenomena like Kondo effect. 

In this paper, we study effect of impurities for strongly coupled system using the gravity dual. We consider an s-wave holographic superconductor where two vector fields $A_\mu$ and $B_\mu$ are introduced in the AdS space: $A_\mu$  is dual to the conserved current in the boundary identified as the weakly-gauged electromagnetic field in the context of superconductors, while the other is massive and dual to impurities. More concretely, we assume that the impurities have another type of charge carriers like a hole, for instance, where the hole number is not necessarily conserved due to the capturing of conduction electrons.  
So we assume that the latter is dual to the massive gauge field $B_\mu$. 
We postulate a gauge invariant coupling term to describe the interaction between the two species.

The model we consider was mentioned in \cite{Hashimoto:2012pb}, where the massive vector field could be introduced as impurities, but it was integrated out assuming that the impurities were infinitely heavy.  If the mass is finite, the field at the asymptotic AdS boundary decays not exponentially but by a power law. Therefore, we do not integrate out the massive field, and solve the model explicitly in order to focus on the dynamics in the presence of the massive vector field. Our result will show that the mass gap of the superconductor disappears due to the interaction.
 
The rest of this paper is organized as follows. In section~\ref{sec:Proca}, we study the model. The action and setups for numerical computations are provided in section~\ref{sec:ProcaModel}. Numerical results are shown in section~\ref{sec:ProcaNumerical}. Some further possibilities of the model are discussed in section~\ref{sec:ProcaNegative}. Comments on normal phase are given in section~\ref{sec:RN}. In section~\ref{sec:Hashimoto}, we reexamine the strategy of \cite{Hashimoto:2012pb}, and discuss relations to ours. We conclude this paper with future perspectives in section~\ref{sec:Conclusion}.

\section{Impurity degrees of freedom by a massive vector field}
\label{sec:Proca}

We consider a holographic superconductor where an extra massive vector field is introduced as impurities. 
 
\subsection{The model}
\label{sec:ProcaModel}

We consider a model where a massive vector field is introduced into the minimal Abelian-Higgs model of the s-wave holographic superconductor \cite{Hartnoll:2008vx}. The action is\footnote{We set the gauge coupling of $A_{\mu}$ as $e=1$.}
\begin{align}
S = \int d^4x \sqrt{-g}
\bigg( &-\frac{1}{4} F_{\mu\nu} F^{\mu\nu} - | \partial_{\mu} \Phi - i A_{\mu} \Phi |^2 - M^2 \Phi^2 \nonumber \\
& - \frac{1}{4} G_{\mu\nu} G^{\mu\nu} - \frac{m^2}{2} B_{\mu} B^{\mu} - \frac{c}{2} F_{\mu\nu} G^{\mu\nu} \bigg), \label{Proca_action}
\end{align}
where $F=dA$ and $G=dB$. The scalar field $\Phi$ is charged only under $A_{\mu}$, and there is no direct coupling between $\Phi$ and $B_{\mu}$. Here $m^2$ is the mass of $B_{\mu}$. It might be possible to consider to generate this mass by some Higgs mechanism, but here we would like to start from the Proca action for simplicity. The scalar mass is chosen to be  $M^2=-2$ for  convenience in analysis.

The action \eqref{Proca_action} has an interaction term between $F_{\mu\nu}$ and $G_{\mu\nu}$. These vector fields are dual to two currents with different dimensionality, and this interaction represents a coupling of the two currents \cite{Hashimoto:2012pb}. One current is conserved fermion number, and $A_{\mu}$ is identified as the gauge field of the weakly-gauged $U(1)$ electromagnetic symmetry in the context of superconductors. The other, $B_{\mu}$, represents impurities. If we allow the difference in anomalous dimensions of the two current operators, the vector field dual to the impurities can be massive and associated with the Proca field.

We work in a limit in which the matter fields do not give back-reactions on the background metric.\footnote{If we recover the gauge coupling $e$ and rescale each field by a factor of $1/e$, then there is a factor $1/e^2$ appearing in front of the right hand side of  \eqref{Proca_action}. Backreactions on the metric are suppressed in the limit $e^2\to\infty$ with the fields fixed. The result takes the same form as \eqref{Proca_action}, where $e=1$.} This limit is called a probe limit. The gravity background we consider is the AdS-Schwarzschild black hole,
\begin{align}
ds^2 = \frac{1}{z^2} \left( -f(z) dt^2 + \frac{dz^2}{f(z)} + dx^2 + dy^2 \right), \quad f(z)=1-z^3, \label{ads_sch_metric}
\end{align}
where we use units in which the AdS radius is unity, and the location of the black hole horizon is at $z=1$. Such parameter fixings are possible thanks to the isometry of the AdS space, and can be confirmed by examining the symmetry of the equations of motion.

The equations of motion of \eqref{Proca_action} are
\begin{align}
& \nabla_{\lambda} F^{\lambda\mu} + c \nabla_{\lambda} G^{\lambda\mu}
- 2 |\Phi|^2 A^{\mu} + i(\Phi^\ast \partial^{\mu} \Phi - \partial^{\mu} \Phi^\ast \Phi) = 0, \label{Proca_A_eom} \\
& \nabla_{\lambda} G^{\lambda\mu} + c \nabla_{\lambda} F^{\lambda\mu} - m^2 B^{\mu} = 0, \label{Proca_B_eom} \\
& \frac{1}{\sqrt{-g}} (\partial_{\mu} - i A_{\mu}) \left( \sqrt{-g} g^{\mu\nu} (\partial_{\nu} - i A_{\nu}) \Phi \right) - M^2 \Phi = 0. \label{Proca_Phi_eom}
\end{align}
The kinetic terms of $A^{\mu}$ and $B^{\mu}$ can be separated. From \eqref{Proca_A_eom} and \eqref{Proca_B_eom}, we obtain
\begin{align}
& (1-c^2) \nabla_{\lambda} F^{\lambda\mu} - 2 |\Phi|^2 A^{\mu} + c \, m^2 B^{\mu}
+ i(\Phi^\ast \partial^{\mu} \Phi - \partial^{\mu} \Phi^\ast \Phi) = 0, \\
& (1-c^2) \nabla_{\lambda} G^{\lambda\mu} - \, m^2 B^{\mu} + 2c |\Phi|^2 A^{\mu}
- ic(\Phi^\ast \partial^{\mu} \Phi - \partial^{\mu} \Phi^\ast \Phi) = 0.
\end{align}

We can realize an superconductor phase where the scalar operator dual to $\Phi$ acquires nonzero condensate $\langle \mathcal{O} \rangle$. The ansatz is $A=A_t(z) dt, B=B_t(z) dt$, and $\Phi=\phi(z)$. The equations of motion become
\begin{align}
& A_t'' - \frac{2 \phi^2 A_t}{(1-c^2) z^2 f} + \frac{c \, \widetilde{m}^2 B_t}{z^2 f} = 0, \label{Proca_At_eom} \\
& B_t'' - \frac{\widetilde{m}^2 B_t}{z^2 f} + \frac{2c \, \phi^2 A_t}{(1-c^2) z^2 f} = 0, \label{Proca_Bt_eom} \\
& \phi'' + \left( \frac{f'}{f} - \frac{2}{z} \right) \phi' + \left( \frac{2}{z^2 f} + \frac{A_t^2}{f^2} \right) \phi = 0. \label{Proca_phi_eom} 
\end{align}
The prime denotes the derivative with respect to $z$: $A_t' = \partial_z A_t$. We find it convenient to define an effective mass of $B_{\mu}$, 
\begin{align}
\widetilde{m}^2 \equiv \frac{m^2}{1-c^2}.
\label{def_mass} 
\end{align}

The system is not well-defined if $c=1$. If we consider a small  deformation from a system without the impurity coupling, then $0\le c<1$ would be reasonable.\footnote{If $c$ is negative, solutions of $B_{\mu}$ flips the sign.}  For the  strong coupling regime, we can take $c>1$. In this case, however,  $A_t$ is tachyonic, and there can be  some instability as we will  discuss in section~\ref{sec:ProcaNegative}. We consider the case that $0\le c < 1$ first.

Data of the boundary theory are extracted from the asymptotic behaviors of the bulk fields near the boundary ($z\to0$). Let us start from $c=0$ case, where $B_{\mu}$ decouples from $A_{\mu}$ and $\phi$. The asymptotic solutions to \eqref{Proca_At_eom} and \eqref{Proca_phi_eom}   take the form, 
\begin{align}
A_t &= \mu - \rho z + \cdots, \label{Proca_At_boundary}\\
\phi &= \phi_1 z + \phi_2 z^2 + \cdots, \label{Proca_phi_boundary}
\end{align}
where $\mu, \, \rho, \phi_i \, (i=1,2)$ are integral constants. Coefficients of higher-order terms are given in terms of these four constants. Here $\mu$ and $\rho$ are the chemical potential and charge density of the $U(1)$ gauge field, respectively, and $\phi_i$ corresponds to the source or condensate of the scalar field. We impose boundary conditions such that the source of the scalar field is zero. We may consider either $\phi_1=0$ or $\phi_2=0$: Since $M^2=-2$ is below the unitarity bound \cite{Klebanov:1999tb}, the conformal dimension given by $\Delta(\Delta-3)=M^2$ has two normalizable solutions $\Delta=1, \, 2$. The boundary conditions $\phi_2=0$ and $\phi_1=0$ correspond to $\Delta=1$ and $\Delta=2$, respectively.

If $c=0$, eq.~\eqref{Proca_Bt_eom} reduces to 
\begin{align}
B_t'' - \frac{m^2 B_t}{z^2 f} = 0. \label{Proca_ProcaEq}
\end{align}
The asymptotic solution to this equation takes the form
\begin{align}
B_t = \beta_{-} z^{a_{-}} + \cdots + ( \beta_{+} + b_{+} \log z ) z^{a_{+}} + \cdots, \quad
a_{\pm} = \frac{1}{2} \left(1 \pm \sqrt{1+ 4 m^2} \right), \label{Proca_Bt_boundary}
\end{align}
where $\beta_{\pm}$ are integral constants: $\beta_{-}$ and $\beta_{+}$ correspond to the source and the condensate, respectively. The logarithmic term may appear only if $a_{+} - a_{-} \in \mathbb{Z}$. The same statement applies later without mentioning explicitly. Solutions exist when the mass is above the BF bound \cite{Breitenlohner:1982jf}, $m^2 \ge -1/4$. If $m^2=0$, the $U(1)$ symmetry of $B_{\mu}$ is restored, and $a_{\pm} = 0, \, 1$.

Now we take into  account the coupling in the equations of motion. The series expansions \eqref{Proca_At_boundary}, \eqref{Proca_phi_boundary} and \eqref{Proca_Bt_boundary} are modified so that the asymptotic solutions accommodate the interaction term. For simplicity, we focus on the case that $\alpha_{\pm}$ defined below are integers. We also impose that the source of $B_t$ is absent: $\beta_{-}=0$. The asymptotic solutions then take the form
\begin{align}
A_t &= \mu - \rho z + \cdots + (a_{+} z^{\alpha_{+}} + \cdots) \log z + \cdots, \label{Proca_integer_At_boundary} \\
B_t &= \beta_{+} z^{\alpha_{+}} + \cdots + (b_{+} z^{\alpha_{+}} + \cdots) \log z + \cdots, \label{Proca_integer_Bt_boundary} \\
\phi &= \phi_1 z + \phi_2 z^2 + \cdots + (\varphi_{+} z^{\alpha_{+} + 2} + \cdots) \log z + \cdots, \label{Proca_integer_phi_boundary}
\end{align}
where
\begin{align}
\alpha_{\pm} = \frac{1}{2} \left(1 \pm \sqrt{1+ 4 \widetilde{m}^2} \right).
\end{align}
Note that the presence of the possible logarithmic terms is due to the impurity coupling. The expansions contain five integral constants $\mu$, $\rho$, $\beta_{+}$, $\phi_1$, and $\phi_2$. We further impose that the source of the scalar is absent: $\phi_i=0 \,  (i=1$ or $2)$. We will later show the asymptotic solutions in the cace that $\widetilde{m}^2 = 2$.\footnote{We obtain $(\alpha_{-}, \alpha_{+}) = (-1,\,2)$ when $\widetilde{m}^2=2$.}

For the asymptotic forms of the solution at the horizon, we require that the gauge fields are zero at the horizon, while the scalar field can have a finite value. The series expansions take the form
\begin{align}
A_t &= A_t^{(1)} (z-1) + A_t^{(2)} (z-1)^2 +\cdots, \\
B_t &= B_t^{(1)} (z-1) + B_t^{(2)} (z-1)^2 +\cdots, \\
\phi &= \phi^{(0)} + \phi^{(1)} (z-1) + \phi^{(2)} (z-1)^2 +\cdots,
\end{align}
where $A_t^{(1)}, \, B_t^{(1)}$ and $\phi^{(0)}$ are the integral constants, and higher-order coefficients are determined in terms of them.

We will numerically solve \eqref{Proca_At_eom}, \eqref{Proca_Bt_eom}, and \eqref{Proca_phi_eom} to compute the condensate. Imposing the boundary conditions, we are left with four undetermined integral constants at the boundary, while there are three at the horizon. Thus we obtain one-parameter family solutions. Each solution is characterized by the temperature $T$ of the system.

It should be noted that $B_t$ can become nontrivial in the bulk due to the coupling term in \eqref{Proca_Bt_eom}, which is nonzero if the scalar condensate is nonzero.  In this case, $B_t$ is inevitable to become nontrivial in the bulk even though the boundary conditions are $B_t|_{z=0, \, z=1}=0$. Eventually, we should obtain spontaneous condensation of $B_t$ coming along with the condensation of the scalar field.

Once we obtain nonzero condensate of the scalar field, we shall examine the properties of this ordered phase by computing the optical conductivity. The ansatz of electric perturbations is $A_x = A_x(z) e^{-i \omega t}$ and $B_x = B_x(z) e^{-i \omega t}$. The equations of motion are
\begin{align}
& A_x'' + \frac{f'}{f} A_x' + \left( \frac{\omega^2}{f^2}  - \frac{2 \phi^2}{(1-c^2) z^2 f} \right) A_x + \frac{c \, \widetilde{m}^2 B_x}{z^2 f} = 0, \label{Proca_Ax_eom} \\
& B_x'' + \frac{f'}{f} B_x' + \left( \frac{\omega^2}{f^2}  - \frac{\widetilde{m}^2}{z^2 f} \right) B_x+ \frac{2 c \, \phi^2 A_x}{(1-c^2) z^2 f} = 0. \label{Proca_Bx_eom}
\end{align}
We impose the ingoing boundary condition at the horizon. The asymptotic  solutions at the horizon hence take the form
\begin{align}
A_x &= f^{-i \omega/3} \left( \widetilde{A}_x^{(0)} + \widetilde{A}_x^{(1)} (z-1) + \widetilde{A}_x^{(2)} (z-1)^2 +\cdots \right), \\
B_x &= f^{-i \omega/3} \left( \widetilde{B}_x^{(0)} + \widetilde{B}_x^{(1)} (z-1) + \widetilde{B}_x^{(2)} (z-1)^2 +\cdots \right).
\end{align}
Higher-order coefficients are determined in terms of the integral constants $\widetilde{A}_x^{(0)}$ and $\widetilde{B}_x^{(0)}$. One of these constants can be fixed thanks to the scaling symmetry of \eqref{Proca_Ax_eom} and \eqref{Proca_Bx_eom} under $(A_x, \, B_x) \to (\lambda A_x, \lambda B_x)$ with some $\lambda$. For instance, we may fix $\widetilde{A}_x^{(0)}=1$. The asymptotic solutions at the boundary in general take the form
\begin{align}
A_x &= A_x^{(-)} z^{\alpha_{-}} + \cdots + A_x^{(0)} + A_x^{(1)} z + \cdots + (\tilde{a}_{0} + \cdots + \tilde{a}_{+} z^{\alpha_{+}} + \cdots) \log z+ \cdots , \label{Proca_Ax_boundary} \\
B_x &= B_x^{(-)} z^{\alpha_{-}} + \cdots +B_x^{(+)} z^{\alpha_{+}} + \cdots + (\tilde{b}_{0} + \cdots + \tilde{b}_{+} z^{\alpha_{+}} + \cdots) \log z + \cdots. \label{Proca_Bx_boundary}
\end{align}
There are four integral constants at the boundary: $A_x^{(0)}, \, A_x^{(1)}, \, B_x^{(-)}$ and $B_x^{(+)}$. Other coefficients can be given in terms of them.

The currents dual to these fluctuations $A_x$ and $B_x$ are computed from the on-shall action in the bulk. The action \eqref{Proca_action} can be separated into the bulk and boundary terms, $S = S_\mathrm{bulk} + S_b$, where the former becomes zero because of the equations of motion, and the latter is used in computing correlation functions under the AdS/CFT prescription. The boundary action is obtained as
\begin{align}
S_b = \frac{1}{2} \int d^3x \left[ A_x \partial_z (A_x + c B_x) + B_x \partial_z (B_x + c A_x) \right]_{z\to0}, \label{Proca_bdryaction}
\end{align}
To compute the on-shell boundary action, we substitute the asymptotic expansions \eqref{Proca_Ax_boundary} and \eqref{Proca_Bx_boundary} into the boundary action, and then carry out the holographic renormalization \cite{deHaro:2000xn,Skenderis:2002wp}. As a result, we obtain a quadratic action for fluctuations $A_x$ and $B_x$. The quadratic action could be computed with general $\widetilde{m}^2$. For simplicity, however, we denote the resultant quadratic action just as $S_b^{(2)}$ here, and will show an explicit computation in the case that $\widetilde{m}^2=2$ in the next section.

The AdS/CFT correspondence tells us that the currents of $A_x$ and $B_x$ are computed from the quadratic boundary action,
\begin{align}
J_x^A = \frac{\delta S_b^{(2)}}{\delta A_x^{(0)}}, \quad
J_x^B = \frac{\delta S_b^{(2)}}{\delta B_x^{(-)}}. \label{Proca_currents}
\end{align}
We call $J_x^A$ as an electric current, and $J_x^B$ as a hole current. 
Using these, we can give the two electric conductivity $\sigma_A$ and $\gamma_B$  associated with two types of charge carriers, 
\begin{align}
\sigma_A(\omega) = \left. \frac{J_x^A}{E_x} \right|_{B_x^{(-)}=0}, \quad
\gamma_B(\omega) = \left. \frac{J_x^B}{E_x} \right|_{B_x^{(-)}=0}, \label{Proca_conductivity}
\end{align}
where the external electric field applied on the system is given by $E_x = -\partial_t A_x^{(0)}$. We numerically solve the equations of motion using the shooting method, and compute the conductivity in the superconductor background.

\subsection{Results}
\label{sec:ProcaNumerical}

For simplicity, we consider the case that $\widetilde{m}^2=2$. We choose the coupling $c=0.5$. The boundary condition of the scalar field is either $\phi_2=0$ or $\phi_1=0$, corresponding to $\Delta=1$ or $\Delta=2$, respectively. In the figures, the orange dotted lines correspond to the case that $c=0$, while the blue real lines correspond to the case of nonzero $c$.

\begin{figure}[t]
\centering
\includegraphics[width=6cm]{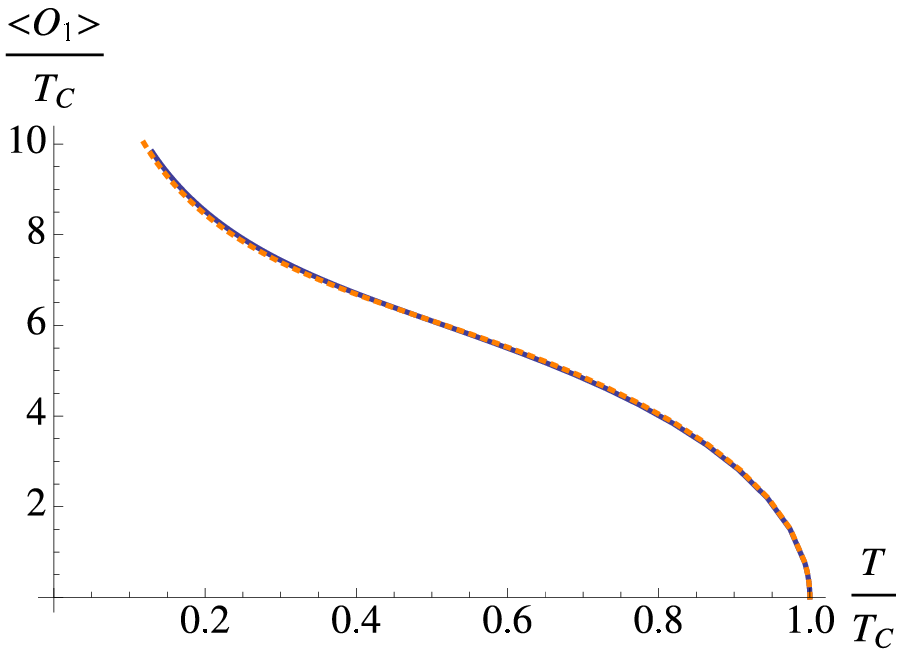}
\includegraphics[width=6cm]{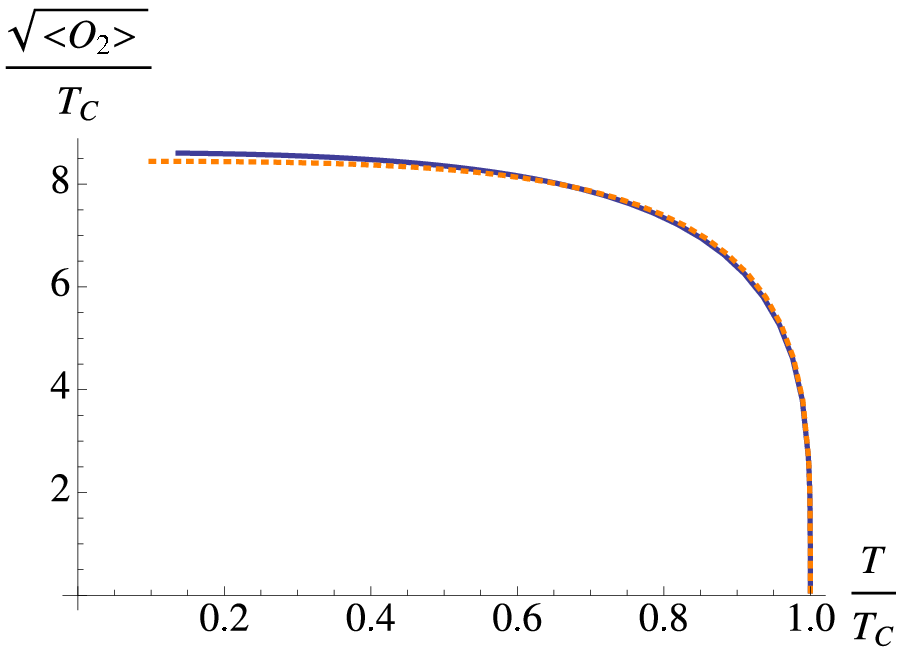}
\caption{The scalar condensate $\langle \mathcal{O} \rangle$ as a function of temperature. The blue real lines are when $\widetilde{m}^2=2, \, c=0.5$. The orange dotted lines reproduce the results in \cite{Hartnoll:2008vx}.}
\label{fig:Proca_condensate_m2c05}
\end{figure}

First, we compute the condensate for superconductor background. As announced, we show the asymptotic solution at the boundary when $\widetilde{m}^2=2$, after $\beta_{-}=0$ is imposed,
\begin{align}
A_t &= \mu -\rho z + \left(  \mu \, \phi_1^2 - c \,\beta_{+} + \frac{2 c^2 \mu \, \phi_1^2 }{3(1-c^2)} \log z \right) z^2 + \cdots, \label{Proca_At_Asymp_m2} \\
B_t &= \left( \beta_{+} - \frac{2 c \mu \, \phi_1^2 }{3(1-c^2)} \log z\right) z^2 + \frac{c \, \phi_1(\rho \,\phi_1 - 2\mu \, \phi_2)}{2(1-c^2)} z^3 + \cdots, \label{Proca_Bt_Asymp_m2} \\
\phi &= \phi_1 z + \phi_2 z^2 - \frac{\mu^2 \phi_1}{2} z^3 + \frac{\phi_1 + 2 \mu \, \rho \, \phi_1 -\mu^2 \phi_2}{6} z^4 + \cdots.\label{Proca_phi_Asymp_m2}
\end{align}
Figure~\ref{fig:Proca_condensate_m2c05} shows the results of scalar condensate. The critical temperature $T_C$ does not change regardless of $c$ because $B_t=0$ at $T=T_C$, where $\phi=0$. Thus we normalize the temperature by $T_C \sim 0.23 \sqrt{\rho}$ in the case that $\Delta=1$, and $T_C \sim 0.12 \sqrt{\rho}$ in the case that $\Delta=2$. We find that the scalar condensate is almost the same as the $c=0$ case, although there can be seen slight deviations in low temperatures.

\begin{figure}[t]
\centering
\includegraphics[width=6cm]{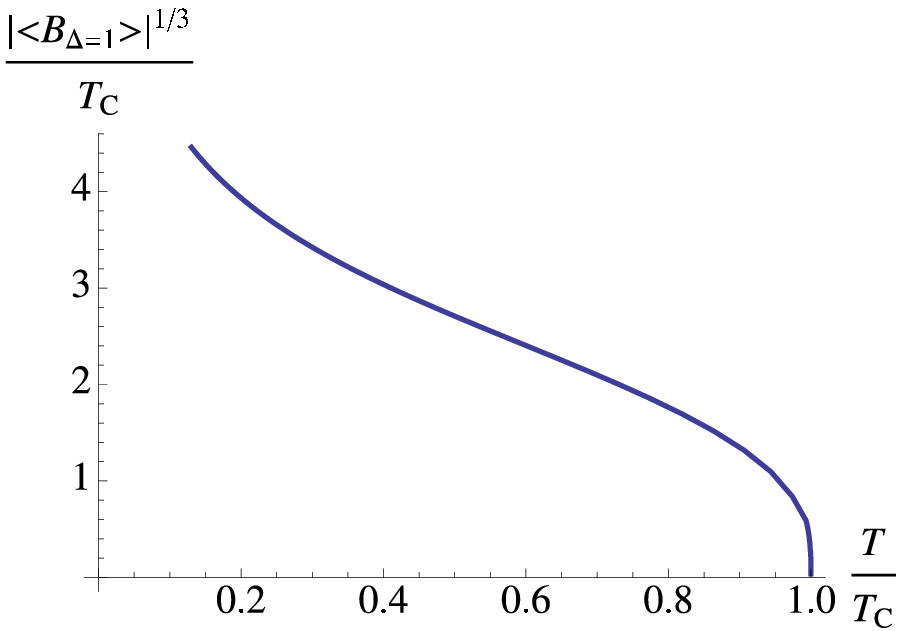}
\includegraphics[width=6cm]{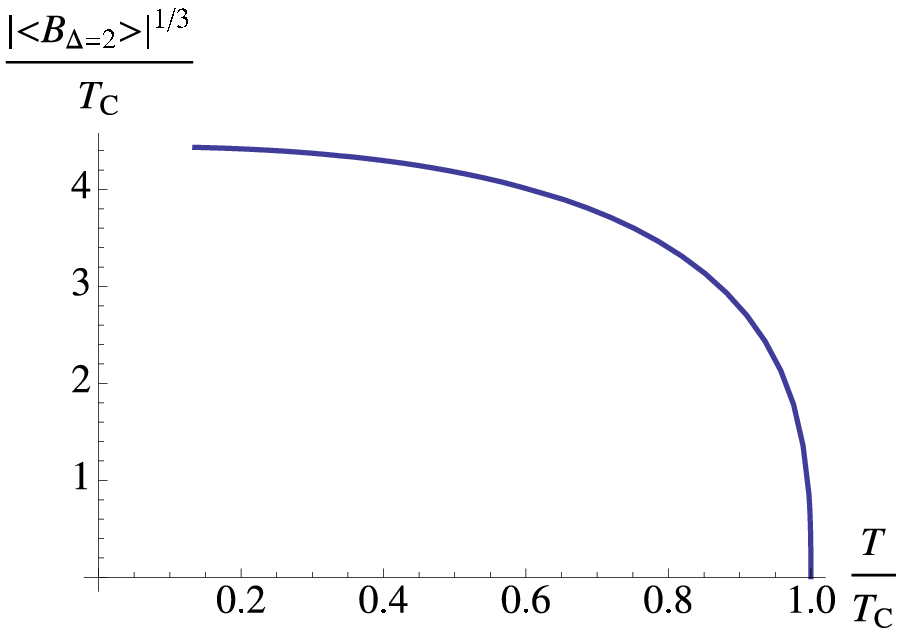}
\caption{The condensate of $B_t$ as a function of temperature when $\widetilde{m}^2=2, \, c=0.5$. In both panels, the condensate goes to zero linearly at $T_C$: $\langle \mathcal{B} \rangle \propto T-T_C$.}
\label{fig:Proca_vevBt_m2c05}
\end{figure}

Figure~\ref{fig:Proca_vevBt_m2c05} shows the condensate of $B_t$, given by $\langle \mathcal{B} \rangle = (1-c^2) \beta_{+} + c \mu \phi_1^2/3$. The absolute value of $\langle \mathcal{B} \rangle$ is plotted with a power of $1/3$.\footnote{The sign of $\langle \mathcal{B} \rangle$ is flipped if the sign of $c$ is negative.} The source of $B_t$ is absent, and the condensation is triggered by $\mu$, but the condensation is present only if the scalar field has nonzero condensate. Figure~\ref{fig:Proca_vevBt_m2c05} tells us that the qualitative behavior of $\langle \mathcal{B} \rangle$ is similar to that of  $\langle \mathcal{O} \rangle$. However, $\langle \mathcal{B} \rangle$ goes to zero linearly at $T \to T_C$. It looks diverging in low temperatures in the case that $\Delta=1$, and backreactions on the gravity background will be important in low temperatures. The condensate $\langle \mathcal{B} \rangle$ becomes nonzero once $c$ is nonzero, and the magnitude becomes larger as we increase $c$.

We compute the conductivity in the superconductor background obtained. First, we prepare the explicit forms of $\sigma(\omega)$ and $\gamma(\omega)$ in the case that $\widetilde{m}^2=2$. The asymptotic solutions of $A_x$ and $B_x$ at the boundary \eqref{Proca_Ax_boundary} and \eqref{Proca_Bx_boundary} are obtained as
\begin{align}
A_x =& - \frac{c B_x^{(-)}}{z} + A_x^{(0)} + \left(A_x^{(1)} - 2 c \, \phi_1^2 B_x^{(-)} \log z \right)z \nonumber \\
& +\Bigg( A_x^{(0)} \left(\phi_1^2 - \frac{\omega^2}{2}\right) - c  (B_x^{(+)} + 2 \phi_1 \phi_2 B_x^{(-)} ) \nonumber \\
& \qquad +\frac{2 c^2 A_x^{(0)} \phi_1^2 +2cB_x^{(-)}(1-c^2(1+4 \phi_1 \phi_2)) }{3(1-c^2)} \log z \Bigg) z^2 + \cdots, \label{Proca_Ax_Asymp_m2} \\
B_x =& \frac{B_x^{(-)}}{z} + \left( \frac{\omega^2}{2}- \frac{c^2\phi_1^2}{1-c^2}\right) B_x^{(-)} z \nonumber \\
&+ \Bigg(B_x^{(+)} - \frac{2 c A_x^{(0)} \phi_1^2 +2B_x^{(-)}(1-c^2(1+4 \phi_1 \phi_2)) }{3(1-c^2)} \log z \Bigg) z^2 + \cdots. \label{Proca_Bx_Asymp_m2}
\end{align}
By making use of them, the currents dual to $A_x$ and $B_x$ are computed as
\begin{align}
J_x^A &= \frac{\delta S_b^{(2)}}{\delta A_x^{(0)}} = A_x^{(1)} + c B_x^{(-)}
\left( \frac{\omega^2}{2} - \frac{(8-5c^2)\phi_1^2}{3(1-c^2)}\right), \label{Proca_JxA} \\
J_x^B &= \frac{\delta S_b^{(2)}}{\delta B_x^{(-)}} = (1-c^2) B_x^{(+)} + c A_x^{(0)}
\left( \frac{\omega^2}{2} - \frac{(8-5c^2)\phi_1^2}{3(1-c^2)}\right). \label{Proca_JxB} 
\end{align}
Thus, two conductivity are given by
\begin{align}
\sigma_A(\omega) &= - \left. \frac{i A_x^{(1)}}{\omega A_x^{(0)}} \right|_{B_x^{(-)}=0}, \label{Proca_ele_conductivity} \\
\gamma_B(\omega) &= - \left.\frac{i}{\omega A_x^{(0)}}
\left((1-c^2) B_x^{(+)} + c A_x^{(0)}
\left( \frac{\omega^2}{2} - \frac{(8-5c^2)\phi_1^2}{3(1-c^2)}\right) \right) \right|_{B_x^{(-)}=0}. \label{Proca_impurity_conductivity}
\end{align}
Notice that the second term in $\gamma_B$ can be found only in the imaginary part.

\begin{figure}[t]
\centering
\includegraphics[width=6cm]{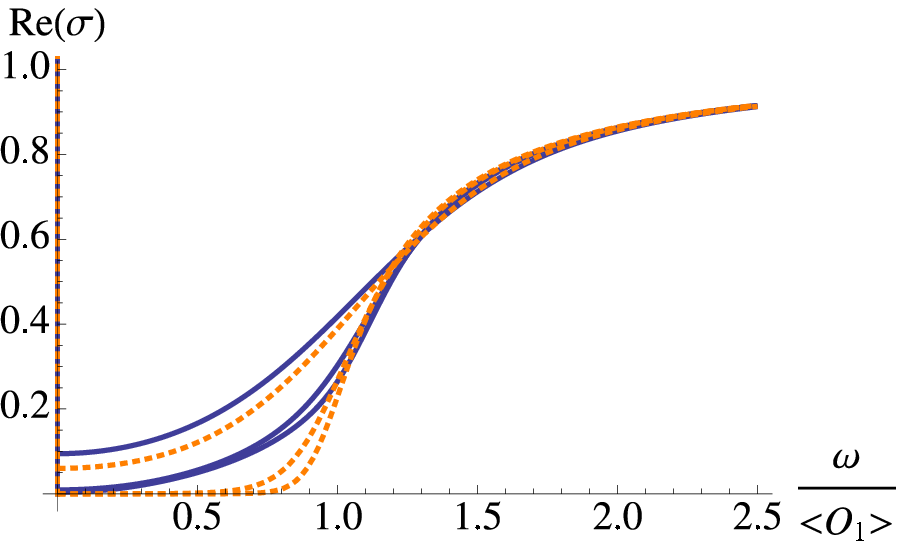}
\includegraphics[width=6cm]{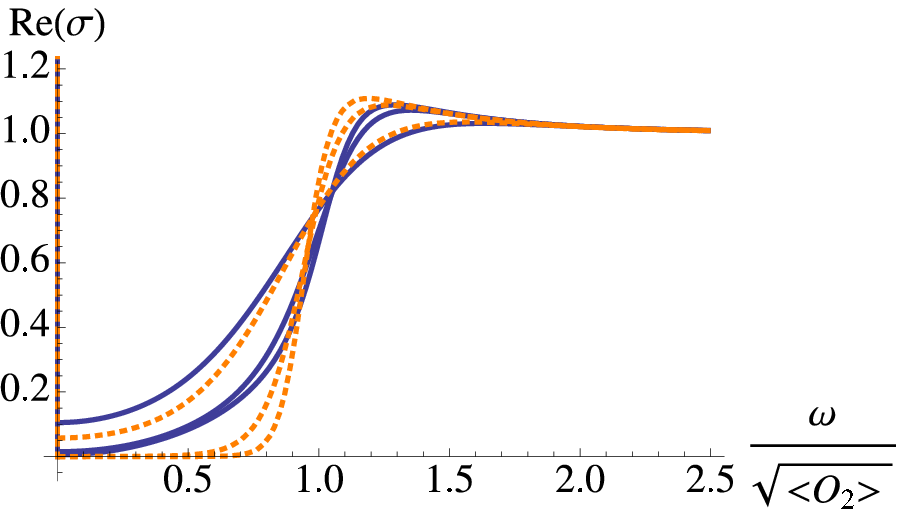}
\caption{The real part of the electric conductivity as a function of $\omega$ normalized by the condensate. The left panel is when $\Delta=1$, and the right panel is when $\Delta=2$. The blue lines correspond to $c=0.5$. The orange lines  go to zero exponentially in low temperatures. However, the blue lines follow power-law behaviors even in low temperatures.}
\label{fig:Proca_gap_Re_m2c05}
\end{figure}

In figure~\ref{fig:Proca_gap_Re_m2c05}, we plot the real part of the electric conductivity as a function of $\omega$ normalized by the scalar condensate. The conductivity is computed for $T/T_C=0.60, \, 0.30, \, 0.20$ when $\Delta=1$, while $T/T_C=0.70, \, 0.35, \, 0.20$ when $\Delta=2$. The rightmost lines correspond to the lowest temperatures.

Let us first recall the interpretation in the case that $c=0$ (orange lines). In the superconducting phase, there is a particular feature that a gap turns up when $T/T_C$ is lowered.  There is no gap until some $T=T_g$ as $\mathrm{Re}(\sigma)|_{\omega \to 0}$ is well above zero. However, when $T/T_C$ is small so that $\mathrm{Re}(\sigma)|_{\omega \to 0}$ is close to zero, the gap quickly shows up when $T<T_g$. This suggests that no quasiparticles are excited up to some $\omega$.

If there are the coupling to the impurities, however, agitation appears in low frequencies as observed in figure~\ref{fig:Proca_gap_Re_m2c05}. The exponential growth of the conductivity in the case of $c=0$ is now replaced with a power-law behavior. The curve in the small frequency region can be fitted with a power function,
\begin{align}
\mathrm{Re} (\sigma) \propto \left(\frac{\omega}{\Lambda_i}\right)^2, \label{Proca_powerlaw}
\end{align}
where $\Lambda_1 \equiv \langle {O_1} \rangle$ and $\Lambda_2 \equiv \sqrt{\langle {O_2} \rangle}$. Thus, this phase is gapless.

\begin{figure}[t]
\centering
\includegraphics[width=6cm]{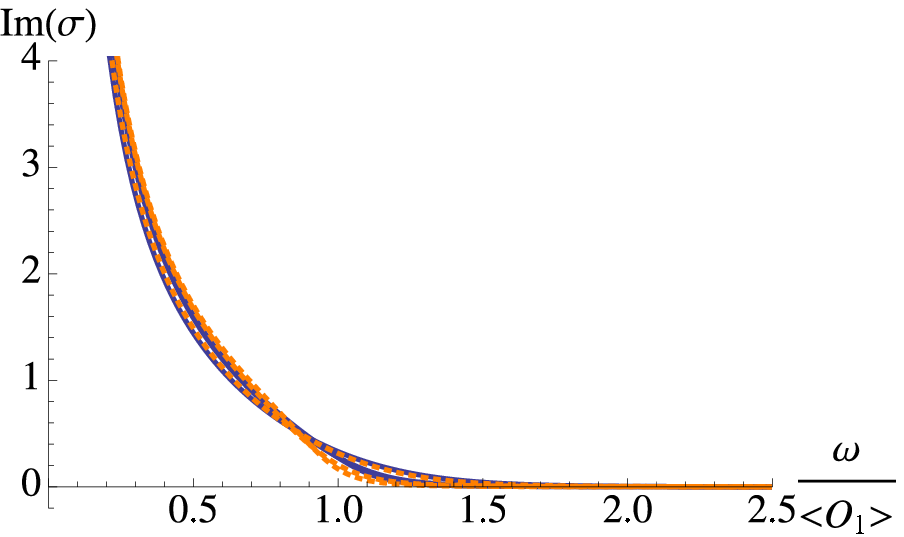}
\includegraphics[width=6cm]{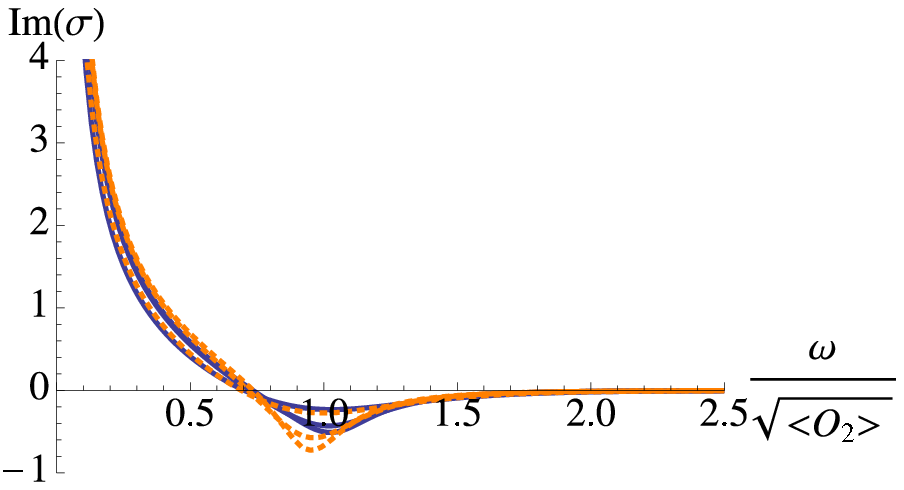}
\caption{The imaginary part of the electric conductivity as a function of $\omega$ normalized by the condensate. The left panel is when $\Delta=1$, and the right panel is when $\Delta=2$.}
\label{fig:Proca_gap_Im_m2c05}
\end{figure}

Figure~\ref{fig:Proca_gap_Im_m2c05} is the imaginary part the conductivity as a function of $\omega$ normalized by the scalar condensate. $\mathrm{Im}(\sigma)$ is diverging as $\omega \to 0$. This behavior corresponds to the presence of the delta function at $\omega=0$ of the real part of the conductivity.
The plot lines slightly shift toward small-$\omega$ direction.

It is expected that the effects of the coupling turn up gradually as the coupling is increased from $c=0$. To discuss this, we compare cases of different $c$. We compute the conductivity for $c=0.2, \, 0.3, \, 0.4, \, 0.5$, by fixing $\widetilde{m}^2$ for simplicity. We have to be careful that this gives the results corresponding to different values of $m^2$ in the action \eqref{Proca_action}. For instance, if $\widetilde{m}^2=2$ is fixed, the computations are for $m^2=1.92, \, 1.82, \, 1.68, \, 1.5$ when $c=0.2, \, 0.3, \, 0.4, \, 0.5$, respectively.\footnote{If we fixed $m^2 =1.5$, the effective masses would be $\widetilde{m}^2= 1.56, \, 1.65, \, 1.79, \, 2$ when $c=0.2, \, 0.3, \, 0.4, \, 0.5$, respectively. We would do computations by employing different asymptotic expansions with respect to different values of $\widetilde{m}^2$.} However, we see that comparing with fixed $\tilde{m}^2$ can give qualitatively reasonable understandings for the behaviors of the conductivity under change of $c$. In particular, we can discuss the presence or absence of the mass gap.

\begin{figure}[t]
\centering
\includegraphics[width=6cm]{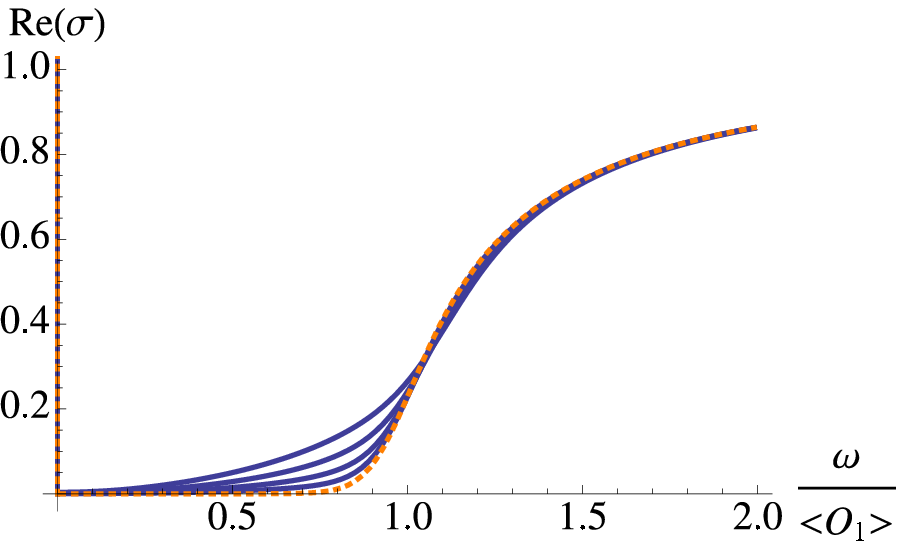}
\includegraphics[width=6cm]{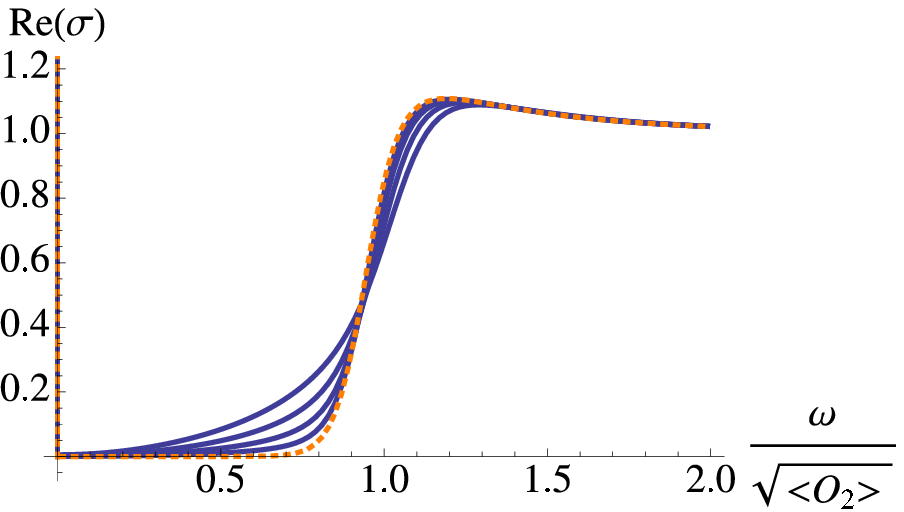}
\caption{Comparisons of the real part of the conductivity for $c=0$ (orange dotted lines) and $c= 0.2, \, 0.3, \, 0.4, \, 0.5$ (blue real lines) computed when $T/T_C=0.20$. The left panel is when $\Delta=1$, and the right panel is when $\Delta=2$. Exponential and power-law behaviors are interpolated by changing $c$.}
\label{fig:Proca_compareRe_T020}
\end{figure}

Comparisons of the real part of the conductivity with respect to different $c$ are given in figure~\ref{fig:Proca_compareRe_T020}. The graphs are computed when $T/T_C=0.20$. We see that, as $c$ is increased, the conductivity gradually changes from exponential to power-law behaviors in the small frequency region. It looks that the transition is continuous, and there would not be a clear phase transition from gapped to ungapped phases under the increase of $c$.

\begin{figure}[t]
\centering
\includegraphics[width=7cm]{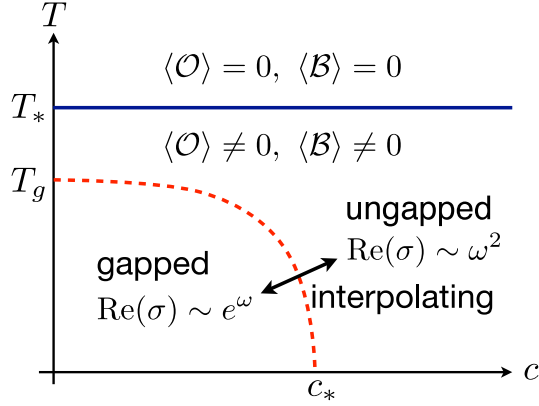}
\caption{A schematic description of the phase diagram expected. Here $\langle \mathcal{B} \rangle = 0$ if $c=0$, while $\langle \mathcal{B} \rangle \neq 0$ once $c\neq0$ and $T<T_C$. Gapped and ungapped phases would be interpolated smoothly. There may be some critical $c=c_\ast$ above which there is no mass gap. Figure~\ref{fig:Proca_compareRe_T020} corresponds to looking at a horizontal slice of this diagram at $T/T_C=0.20$.}
\label{fig:dopingdiagram}
\end{figure}

In figure~\ref{fig:dopingdiagram}, we would like to present a schematic description of the phase diagram of our model under the change of $c$. The strength of the coupling $c$ might be interpreted as the density of impurities. We expect that the regions of the gapped and ungapped phases are smoothly connected under the change of the density. The boarder would not be given by a clear phase transition. When $c=0$, the gap starts to show up at about some $T_g$. As $c$ increases, the temperature for the appearance of the gap looks to be lowered. There may be some critical $c=c_{\ast}$ above which there is no gapped phase even at $T=0$, or $c_{\ast}$ may be located at $c \to 1$. It will be interesting to look for such a quantum critical point. However, to discuss the zero temperature limit, we will need to carry out precise analysis by including back reactions on the gravity backgrounds \cite{Horowitz:2009ij}.

We may consider to vary the mass $\widetilde{m}^2$. When this is heavier, we find that the effects on the conductivity is smaller. Technically, this would be because $\alpha_{+}$ is larger and $B_{\mu}$ decays faster at the boundary. Hence the effects on $A_x$ from $B_x$ would be smaller at the boundary. If the mass is heavier, the massive vector field becomes more non-dynamical, and thus the effects might be harder to appear.

\begin{figure}[t]
\centering
\includegraphics[width=6cm]{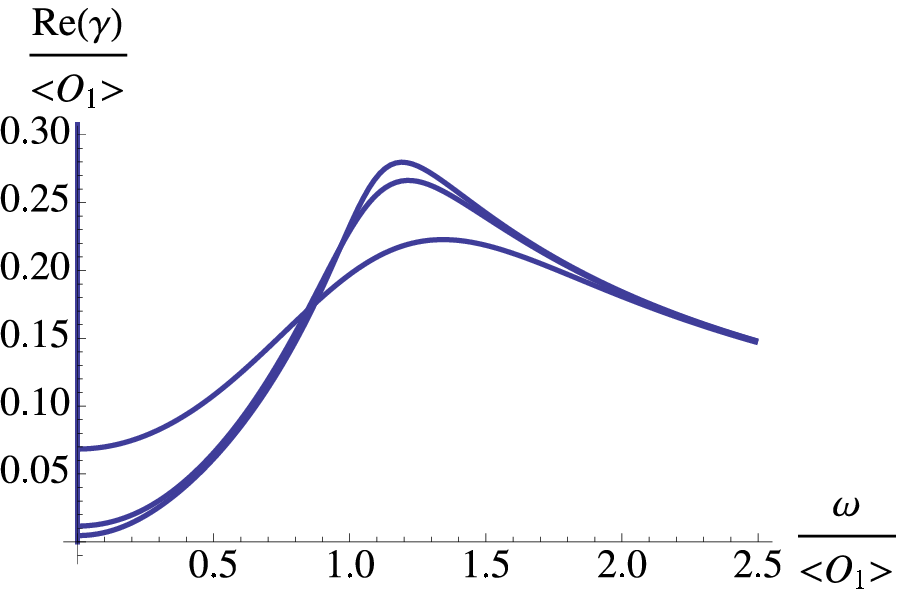}
\includegraphics[width=6cm]{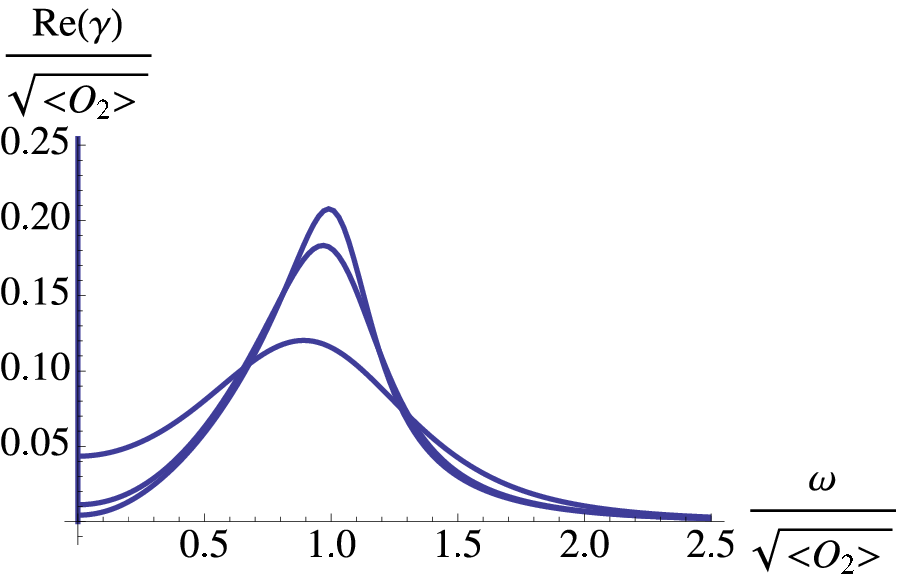}
\caption{The real part of $\gamma$ as a function of $\omega$ normalized by the condensate. The left panel is when $\Delta=1$, and the right panel is when $\Delta=2$. Larger magnitude lines correspond to lower temperatures.}
\label{fig:Proca_impurity_Re_m2c05}
\end{figure}

\begin{figure}[t]
\centering
\includegraphics[width=6cm]{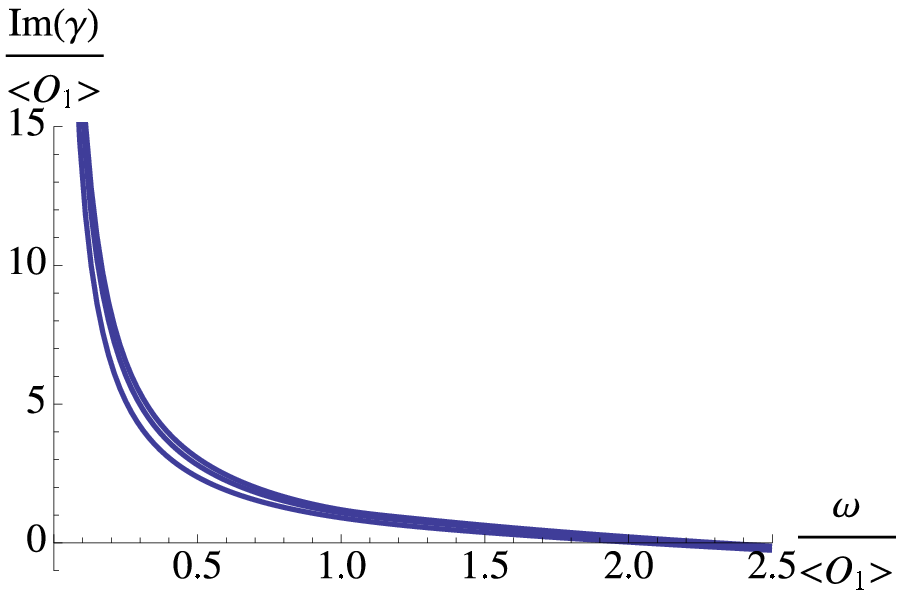}
\includegraphics[width=6cm]{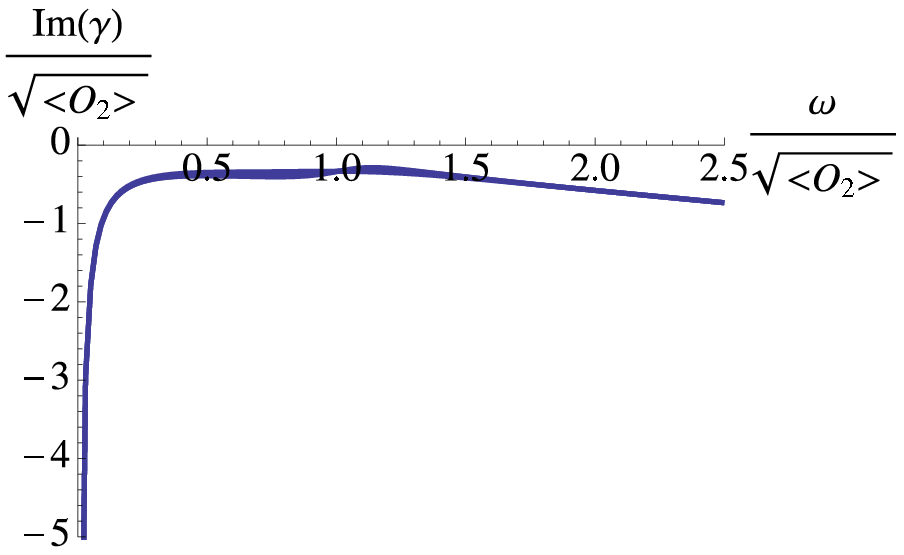}
\caption{The imaginary parts of $\gamma$  as a function of $\omega$ normalized by the condensate. The left panel is when $\Delta=1$, and the right panel is when $\Delta=2$. Larger magnitude lines correspond to lower temperatures.}
\label{fig:Proca_impurity_Im_m2c05}
\end{figure}

Let us come back to the case that $\widetilde{m}^2=2$ and $c=0.5$. The results of $\gamma(\omega)$ are given in figures~\ref{fig:Proca_impurity_Re_m2c05} and~\ref{fig:Proca_impurity_Im_m2c05}, corresponding to the real and imaginary parts, respectively. The conductivity is zero at $T=T_C$ since $B_x=0$ in the normal phase, although we do not plot $T=T_C$ case in the figures. As the temperature is decreased from $T=T_C$, the conductivity starts to grow. There cannot be seen a gap in the small frequency region of the real part as in the case of the electric conductivity. Hence, $\mathrm{Re}(\gamma)$ can be fitted with a power-law function,
\begin{align}
\frac{\mathrm{Re}(\gamma)}{\Lambda_i} - R_0 \propto \left(\frac{\omega}{\Lambda_i}\right)^2,
\end{align}
where $\Lambda_1 \equiv \langle \mathcal{O}_1 \rangle$ and $\Lambda_2 \equiv \sqrt{\langle \mathcal{O}_2 \rangle}$, and $R_0 \equiv \lim_{\omega \to 0} \mathrm{Re}(\gamma)/\Lambda_i$ is an intercept. In figure~\ref{fig:Proca_impurity_Re_m2c05}, a peak can be found in $\mathrm{Re}(\gamma)$ at $w_1 \sim 1.2$ when $\Delta=1$, and $w_2 \sim 1$ when $\Delta=2$. The peak becomes more apparent as the temperature is lowered, and it is expected that $\mathrm{Re}(\gamma)/\Lambda_i$ converges to a finite height curve in the low temperature limit. The nonzero slope in the large-$\omega$ region of $\mathrm{Im}(\gamma)$ is due to the direct contribution of the electric field.

\begin{figure}[t]
\centering
\includegraphics[width=6cm]{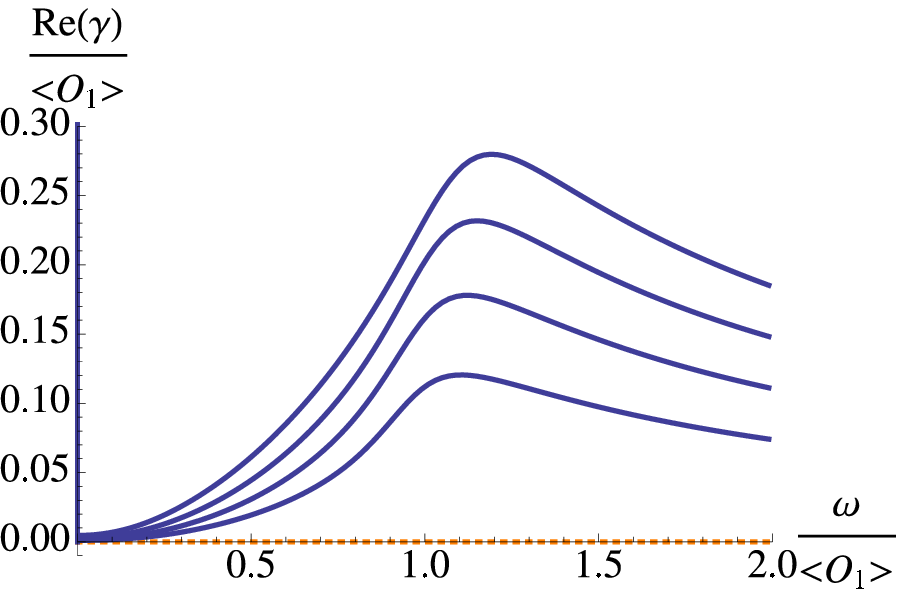}
\includegraphics[width=6cm]{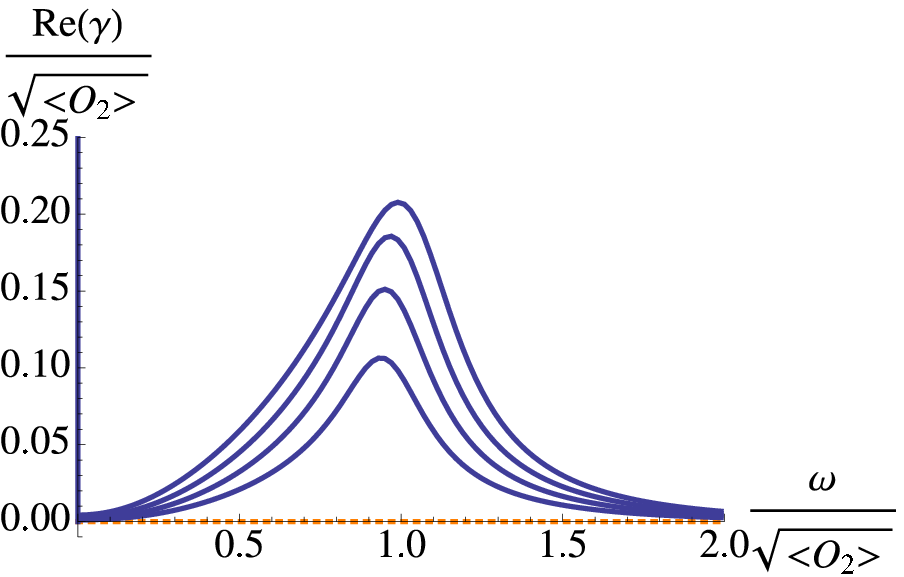}
\caption{Comparisons of the real part of $\gamma$ for $c=0$ (orange dotted lines) and $c= 0.2, \, 0.3, \, 0.4, \, 0.5$ (blue real lines) computed when $T/T_C=0.20$. The left panel is when $\Delta=1$, and the right panel is when $\Delta=2$. Lines with larger magnitude correspond to larger $c$.}
\label{fig:Proca_impuritycompRe_T020}
\end{figure}

Similar to the case of the electric conductivity, we compare the dependence of $\gamma$ on $c$. The results for the real part when $T/T_C=0.20$ are given in figure~\ref{fig:Proca_impuritycompRe_T020}. Since $B_\mu$ is decoupled from $A_\mu$ when $c=0$, $\gamma$ is not induced by the electric perturbations in this limit. From this figure, it can be seen that the position of the peak slightly moves to larger $\omega$ as $c$ is increased. The magnitude of the peak looks to become larger as $c$ is increased. However, since $\mathrm{Re}(\gamma)$ is proportional to $(1-c^2)$, the magnitude of $\gamma$ would be suppressed if $c$ is sufficiently large. It is expected in the heavy mass limit of $\widetilde{m}^2$ that $B_\mu$ would be hard to be excited, and this would mean suppression of $\gamma$ in the $c\to1$ limit.

\subsection{The possibility of negative effective mass}
\label{sec:ProcaNegative}

In the previous section, we considered the case that $0 \le c <1$ and $\widetilde{m}^2=2$. However, since the BF bound for the massive vector field is $m^2=-1/4$, it can be possible that the effective mass of $B_{\mu}$, $\widetilde{m}^2 = m^2/(1-c^2)$, is negative. One possibility is that the system is in strong coupling: $c>1$ with $m^2>0$. The other possibility is that the mass squared is negative $m^2<0$ but $0<c<1$. In the latter case, we obtain $|m^2| < |\widetilde{m}^2|$. Therefore, $m^2$ satisfies the BF and unitarity bounds, $-1/4 < m^2 < 3/4$, if $\widetilde{m}^2$ does. We discuss these options.

First, we consider the possibility if the system is in strong coupling. In this case, we achieve $1-c^2<0$. Hence, the effective mass squared of $A_t$ in \eqref{Proca_At_eom} becomes negative. It is therefore concerned that the effective mass function of $A_t$ may violate the BF bound at some point where $\phi$ is large, although $\widetilde{m}^2$ for $B_t$ is above the BF bound. In fact, numerical examinations in the probe limit tell us that $A_t$ can be badly tachyonic. Therefore, the assumption of strong coupling seems unreasonable.

\begin{figure}[t]
\centering
\includegraphics[width=6cm]{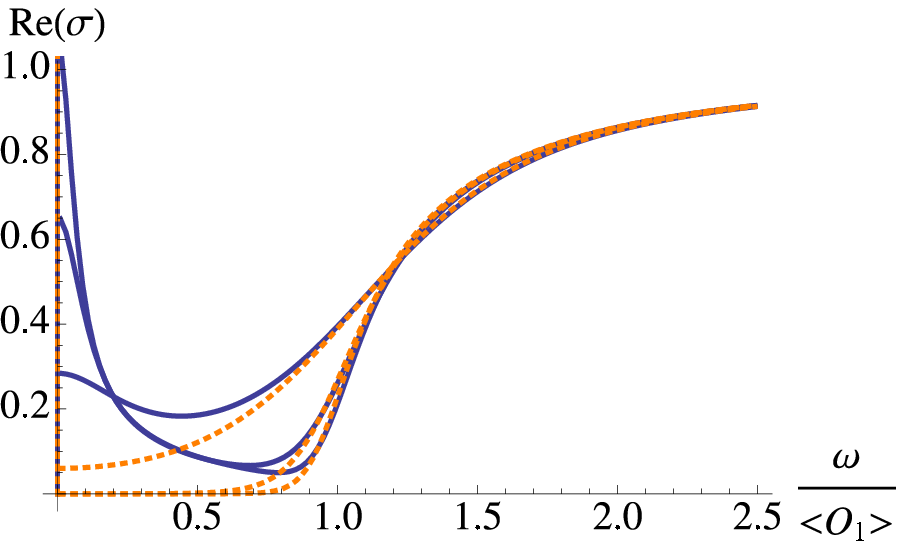}
\includegraphics[width=6cm]{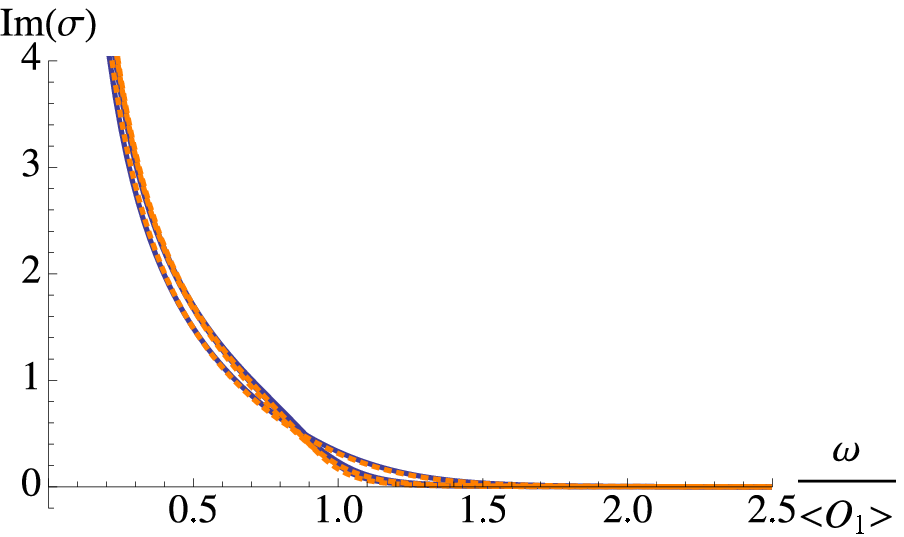}
\caption{The conductivity of $\Delta=1$ case when $\widetilde{m}^2=-1/4$. The blue real lines correspond to this negative effective mass case with $c=0.2$, while the orange dotted lines are when $c=0$. The real part of the conductivity increases near $\omega=0$, but there is a delta function at $\omega=0$.}
\label{fig:negative_gap_O1_mn25c02}
\end{figure}

Second, we discuss the case that $m^2<0$ but $0<c<1$. The scalar condensate is almost the same as that found in figure~\ref{fig:Proca_condensate_m2c05}. Moreover, $\langle \mathcal{B} \rangle$ goes to zero linearly at $T \to T_C$ as well, although the slope  can be different from the previous case.  The conductivity, however, behaves strangely in this case. Figure~\ref{fig:negative_gap_O1_mn25c02} shows the electric conductivity when $\widetilde{m}^2=-1/4$ and $c=0.2$. Curiously, the real part grows up  as  $\omega$ decreases. This may look similar to the presence of the Drude peak. However, as is suggested by the imaginary part, there is a delta function at $\omega=0$ of the real part. It is certain that this rise in small frequency is due to the coupling to $B_x$. In this case of $c=0.2$, the mass gap looks disappearing owing to the uplift of the conductivity. It will be interesting if relevance of these behaviors to real-world condensed matter systems are understood.

\section{Without scalar field: RN-AdS black hole background}
\label{sec:RN}

We have considered the addition of the massive vector field in superconductor phase. However, it will be interesting if we can observe some effects in normal phase. We try a case that the gravity background is the RN-AdS black hole, where the backreaction of the $U(1)$ gauge field is included. Since the massive vector field is not charged, we do not turn on $B_t$, and hence the gravity background is not affected by $B_{\mu}$. Here we discuss that spontaneous induction of $B_{\mu}$ do not occur. This suggests that simply using the RN-AdS black hole might be too naive. It would be interesting to consider more general black hole backgrounds where $B_{\mu}$ can back-react on the gravity backgrounds.

Let us consider a model that contains a $U(1)$ gauge field and a massive vector field but no scalar field. The action is
\begin{align}
S = \frac{1}{2 \kappa^2}\int d^4x \sqrt{-g}
&\left( R + \Lambda -\frac{1}{4} F_{\mu\nu} F^{\mu\nu}  - \frac{1}{4} G_{\mu\nu} G^{\mu\nu} 
- \frac{m^2}{2} B_{\mu} B^{\mu} - \frac{c}{2} F_{\mu\nu} G^{\mu\nu} \right).
\label{RN_action}
\end{align}
By defining the effective mass as $\widetilde{m}^2 \equiv m^2/(1-c^2)$, we obtain the equations of motion of the matter fields as
\begin{align}
\nabla_{\lambda} F^{\lambda\mu} + c \, \widetilde{m}^2 B^{\mu} &= 0, \label{RN_At_eom} \\
\nabla_{\lambda} G^{\lambda\mu} - \widetilde{m}^2 B^{\mu} &= 0. \label{RN_Bt_eom}
\end{align}
These should be accompanied with the Einstein equation.

We try the the Reissner-Nordstr\"{o}m AdS (RN-AdS) black hole as the gravity background,
\begin{align}
ds^2 &= \frac{1}{z^2} \left( -f(z) dt^2 + \frac{dz^2}{f(z)} + dx^2 + dy^2 \right), \\
f(z) &= 1 + (\mu/\gamma)^2 z^4 - (1 + (\mu/\gamma)^2) z^3, \\
A_t &= \mu(1-z),
\label{RN_AdS_metric}
\end{align}
where $\mu$ is the chemical potential, and $\mu/\gamma$ gives the electric charge density of the black hole. The location of the outer horizon is at $z=1$. The Einstein equation can be solved with $B_{\mu}=0$. Note that $B_{\mu}$ is decoupled in \eqref{RN_Bt_eom}. Without source term, $B_{\mu}=0$ can be the solution.

Let us consider the optical conductivity of $A_x$ in this background. The ansatz of electric perturbation is $A_x = A_x(z) e^{-i \omega t}, \, B_x = B_x(z) e^{-i \omega t}$, and $g_{tx} = g_{tx}(z) e^{-i \omega t}$. The linearized equations of motion for the gauge fields are
\begin{align}
& A_x'' + \frac{f'}{f} A_x' + \frac{\omega^2}{f^2} A_x + \frac{z^2 A_t'}{f} \left( g_{tx}' + \frac{2}{z} g_{tx} \right) + \frac{c \, \widetilde{m}^2}{z^2 f} B_x = 0, \label{RN_Ax_eom} \\
& B_x'' + \frac{f'}{f} B_x' + \left( \frac{\omega^2}{f^2} - \frac{\widetilde{m}^2}{z^2 f} \right) B_x = 0, \label{RN_Bx_eom}
\end{align}
The linearized Einstein equation gives a first-order equation of $g_{tx}$,
\begin{align}
g_{tx}' + \frac{2}{z} g_{tx} + \frac{4 A_t'}{\gamma^2} \left( A_x + c B_x \right) = 0. \label{RN_gtx_eom}
\end{align}
We can eliminate $g_{tx}$ from \eqref{RN_Ax_eom} by using \eqref{RN_gtx_eom}. We obtain 
\begin{align}
A_x'' + \frac{f'}{f} A_x' + \left( \frac{\omega^2}{f^2} - \frac{4 z^2 (A_t')^2}{\gamma^2 f} \right) A_x + c \left( \frac{\widetilde{m}^2}{z^2 f} - \frac{4 z^2 (A_t')^2}{\gamma^2 f} \right) B_x = 0. \label{RN_Ax_eom2}
\end{align}
At the horizon, we impose the ingoing boundary condition such that $A_x$ and $B_x$ are proportional to $f^{-i\omega/(3-\mu^2/\gamma^2)}$. Unfortunately, \eqref{RN_Bx_eom} is decoupled from $A_x$, and the solution is $B_x=0$. Thus, there is no contribution of $B_x$.

\section{Comments on adding impurities by bulk source terms}
\label{sec:Hashimoto}

In previous sections, we considered a model where there is a dynamical massive vector field. On the other hand, an idea on introducing impurities in holography non-dynamically was proposed in \cite{Hashimoto:2012pb}. Since our model is related to theirs, we would like to discuss \cite{Hashimoto:2012pb} in this section.

The proposal is that one can introduce impurities by adding a bulk source term, $\int d^4x A_{\mu} J^{\mu}$, where the bulk $U(1)$ source $J^{\mu}$ would correspond to the impurities. Lagrangians with this bulk source term is treated as starting places for analyzing impure systems. This coupling is considered in the s-wave holographic superconductor. The action is
\begin{align}
S = \int d^4x \sqrt{-g} \left( -\frac{1}{4} F_{\mu\nu} F^{\mu\nu} - | \partial_{\mu} \Phi - i A_{\mu} \Phi |^2 - M^2 \Phi^2 +  A_{\mu} J^{\mu} \right), \label{hashimoto_action}
\end{align}
where $M^2=-2$. The probe limit is assumed, and the background is the AdS-Schwarzschild black hole \eqref{ads_sch_metric}. Here we reexamine this model. In particular, we shall understand how the effects of $J^{\mu}$ appear more precisely than the results in \cite{Hashimoto:2012pb}.

Taking \eqref{hashimoto_action} as the starting place may be regarded as an economical simplification in analyzing impurity systems. In giving \eqref{hashimoto_action}, it is assumed that the impurities which would have been introduced in underlying theories are somehow replaced with $J^{\mu}$. We may consider many possible Lagrangians which might reduce to the form of \eqref{hashimoto_action}. For instance, the massive vector model that we focus on in this paper can be found in \cite{Hashimoto:2012pb}, where the massive vector field is interpreted as impurities. However, it might be basically hard to derive $J^{\mu}$ by integrating out the impurity fields explicitly. Therefore, one would start from an effective theory where the bulk source term is introduced as in \eqref{hashimoto_action}. However, the form of $J^{\mu}$ should be arbitrary, and therefore needs to be chosen by hand. Some reasonable $J^{\mu}$ may be assumed.

An example considered in \cite{Hashimoto:2012pb} is to add impurity density by turning on $J^t$, while other components of $J^{\mu}$ are zero. Using the ansatz $A=A_t(z) dt$ and $\Phi=\phi(z)$, we obtain the equations of motion for scalar condensate,
\begin{align}
& A_t'' - \frac{2 \phi^2 A_t}{z^2 f}  = \frac{J^t}{z^4}, \label{hashimoto_At_eom} \\
& \phi'' + \left( \frac{f'}{f} - \frac{2}{z} \right) \phi' + \left( \frac{2}{z^2 f} + \frac{A_t^2}{f^2} \right) \phi = 0. \label{hashimoto_phi_eom} 
\end{align}
Since $J^t$ needs to be chosen by hand, $J^t=\tilde{c} \, z^6$ is adopted. This is devised such that $J^t$ decays fast enough at the boundary. However, how to choose $J^t$ is irrelevant to dynamics of the system. In particular, in this example $J^t \neq 0$ at $T=T_C$. This is different from what we observed in section~\ref{sec:Proca}, where $B_{\mu}$ is zero at $T=T_C$, and nontrivial when $T<T_C$.

Since $J^x$ is not present, the equation of motion for the electric perturbation $A_x = A_x(z) e^{-i \omega t}$ is as usual,
\begin{align}
A_x'' + \frac{f'}{f} A_x' + \left( \frac{\omega^2}{f^2} - \frac{2 \phi^2}{z^2 f} \right) A_x= 0.  \label{hashimoto_Ax_eom}
\end{align}
It is expected that the impurity density $J^t$ will give some effects on $\phi$ through $A_t$, and the conductivity will receive contributions of $J^t$ through $\phi$. Results obtained in \cite{Hashimoto:2012pb} deduce that, compared at some given temperature, the real part of the conductivity increases in the presence of $J^t$.\footnote{A positive $\tilde{c}$ was used in their computation. To be precise, the signs of $\tilde{c}$ and chemical potential $\mu$ are the same. Conventions are such that the chemical potential $\mu$ is also positive.} This would sound similar to our results. Hence, we would like to find out the grounds that lead us to this observation.

The case of nontrivial $J^y$ is also considered in \cite{Hashimoto:2012pb}. The equations of motion for the superconductor background need non-zero $A_y$-component:
\begin{align}
A_y'' + \frac{f'}{f} A_y' - \frac{2 \phi^2 A_y}{z^2 f} + \frac{J^y}{z^4 f} = 0,
\end{align}
together with equations for $A_t$ and $\phi$. Here $J^t=0$. However, the conductivity equation takes the same form as  \eqref{hashimoto_Ax_eom}. The upshot is that the results of the conductivity are qualitatively the same as the case of the impurity density $J^t$. For this reason, we discuss only the case of the impurity density.

We compute the scalar condensate and critical temperatures for superconductivity for various $\tilde{c}$. We use the same impurity density as in \cite{Hashimoto:2012pb}: $J^t=\tilde{c} \, z^6$. Note that the relative signs between $\tilde{c}$ and $\mu$ are important, and $\mu$ is positive in our convention. We find that the critical temperature $T_C$ shifts depending on the magnitude of $\tilde{c}$. Results are given in table~\ref{table:hi_Tc}. For this reason, it is convenient to normalize temperatures by the critical temperature when $\tilde{c}=0$, which we define as $T_{\ast}$. We see that $T_C$ decreases if $\tilde{c}>0$, while increases if $\tilde{c}<0$. The shifts are larger when $\Delta=1$ than when $\Delta=2$.

\begin{figure}[t]
\centering
\includegraphics[width=6cm]{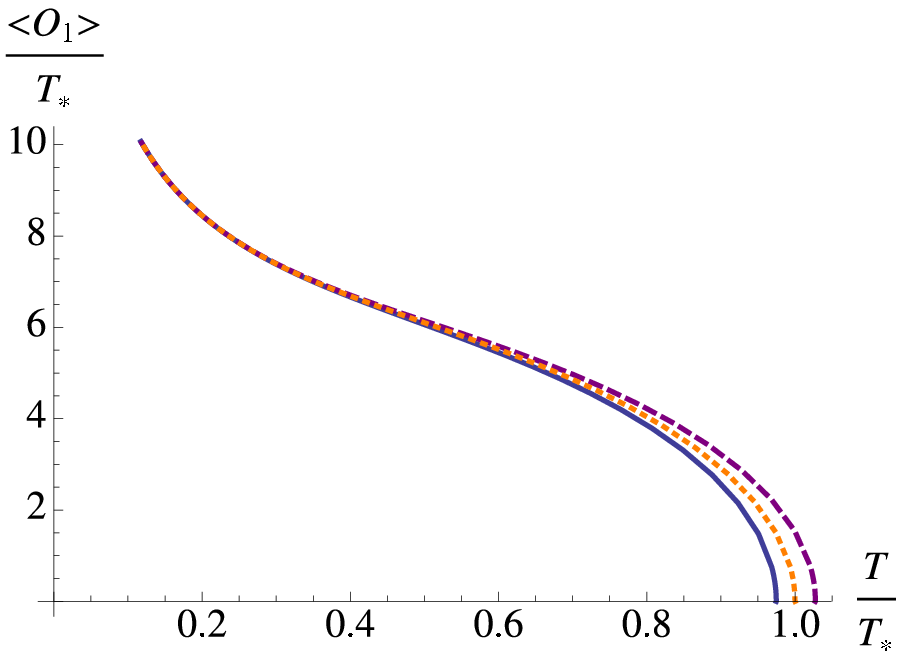}
\includegraphics[width=6cm]{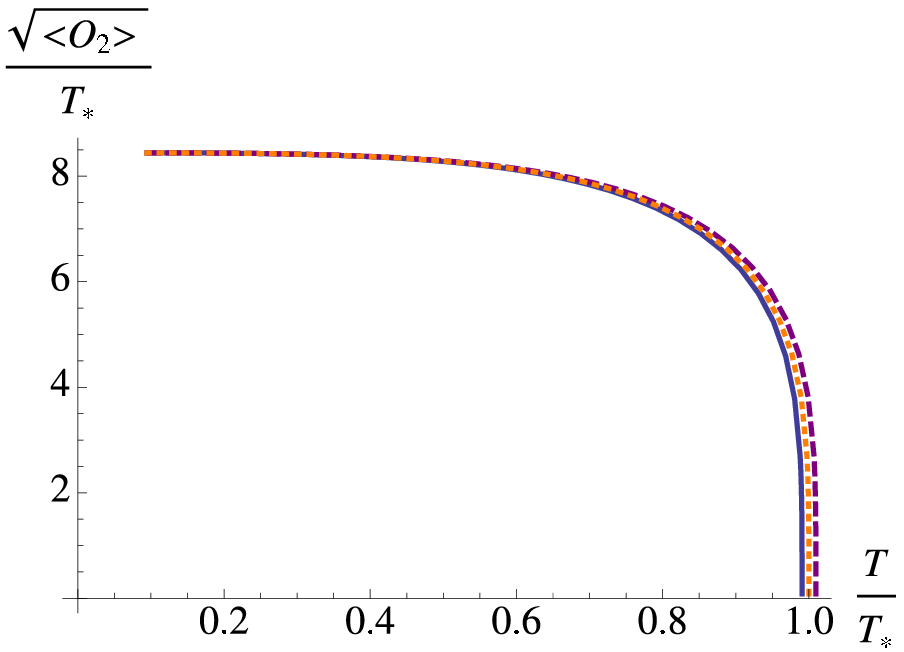}
\caption{The scalar condensate for $\Delta = 1$ (left panel) and $\Delta = 2$ (right panel). The blue real lines are when $\tilde{c}=0.5$, the orange dotted lines are when $\tilde{c}=0$, and the purple dashed lines are when $\tilde{c}=-0.5$. The temperature and condensate are measured in units of $T_{\ast}$. The critical temperatures are different for different values of $\tilde{c}$.}
\label{fig:hi_condensate}
\end{figure}

Figure~\ref{fig:hi_condensate} compares the scalar condensate when $\tilde{c}=0, \, \pm0.5$. Effects of $J^t$ are recognized only near $T_{\ast}$, and disappear when $T/T_{\ast}$ is small. It would be reasonable to think that the addition of impurities change the critical temperatures for the superconductivity. The reason why such shifts of the critical temperature occur seems to be because $J^t \neq 0$ at $T=T_C$, where it looks that the configuration of $A_t$ at $\phi \to 0$ is affected by $J^t$. This may be interpreted such that external operations are applied on the system through $J^t$.

\begin{table}[t]
\begin{center}
\begin{tabular}{c|ccccccc}
\hline
$\tilde{c}$ & $-1$ & $-0.5$ & $-0.1$ & $0$ & $0.1$ & $0.5$ & $1$ \\ \hline
$T_{C}/T_{\ast}|_{\Delta=1}$ & $1.058$ & $1.028$ & $1.005$ & $1$ & $0.995$ & $0.975$ & $0.951$ \\
$T_{C}/T_{\ast}|_{\Delta=2}$ & $1.020$ & $1.010$ & $1.002$ & $1$ & $0.998$ & $0.991$ & $0.982$ \\
\hline
\end{tabular}
\caption{Difference of the critical temperatures from the $\tilde{c}=0$ case. Here $T_{\ast}$ is the critical temperature when $\tilde{c}=0$.}
\label{table:hi_Tc}
\end{center}
\end{table}

\begin{figure}[t]
\centering
\includegraphics[width=6cm]{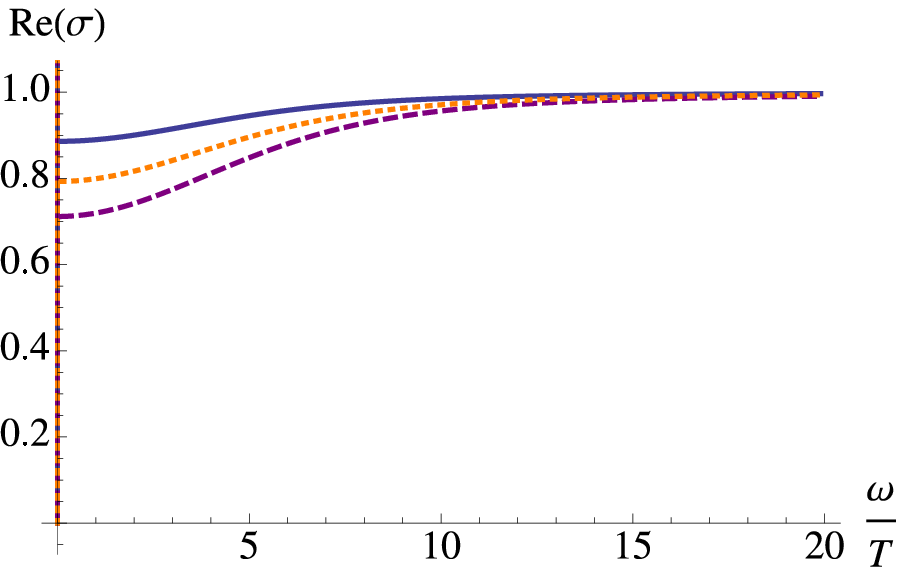}
\includegraphics[width=6cm]{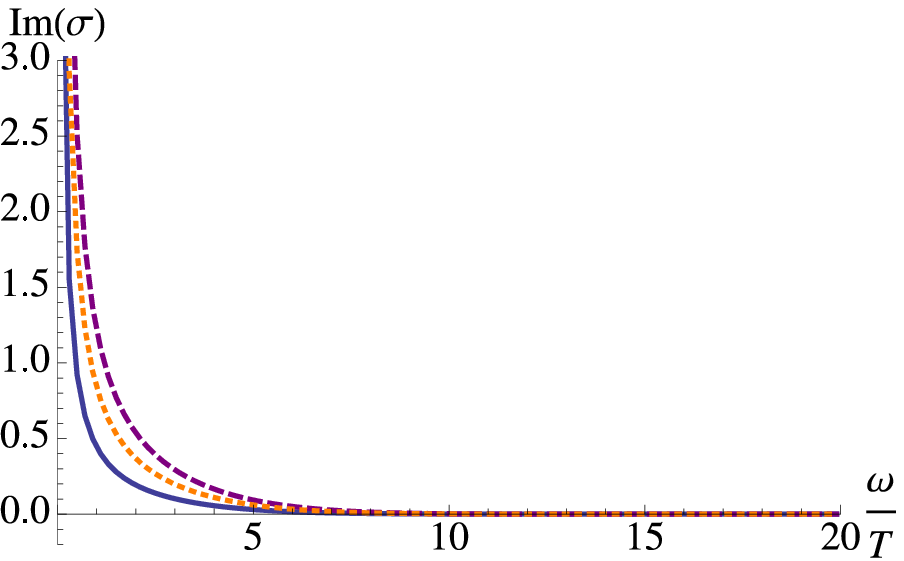}
\caption{The conductivity when $\Delta=1$ and $T/T_{\ast}=0.95$. The blue real lines are when $\tilde{c}=0.5$, the orange dotted lines are when $\tilde{c}=0$, and the purple dashed lines are when $\tilde{c}=-0.5$.}
\label{fig:hi_sigmaO1}
\end{figure}

\begin{figure}[t]
\centering
\includegraphics[width=6cm]{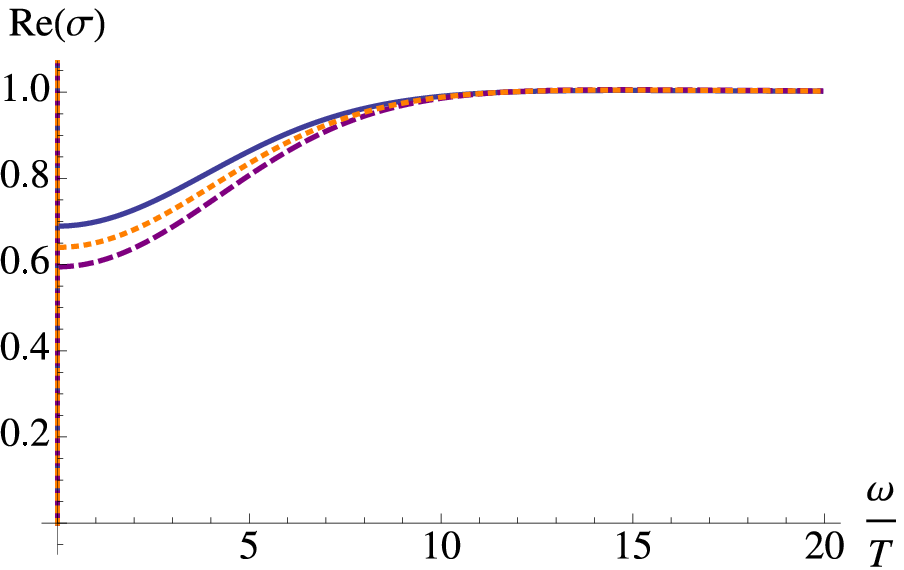}
\includegraphics[width=6cm]{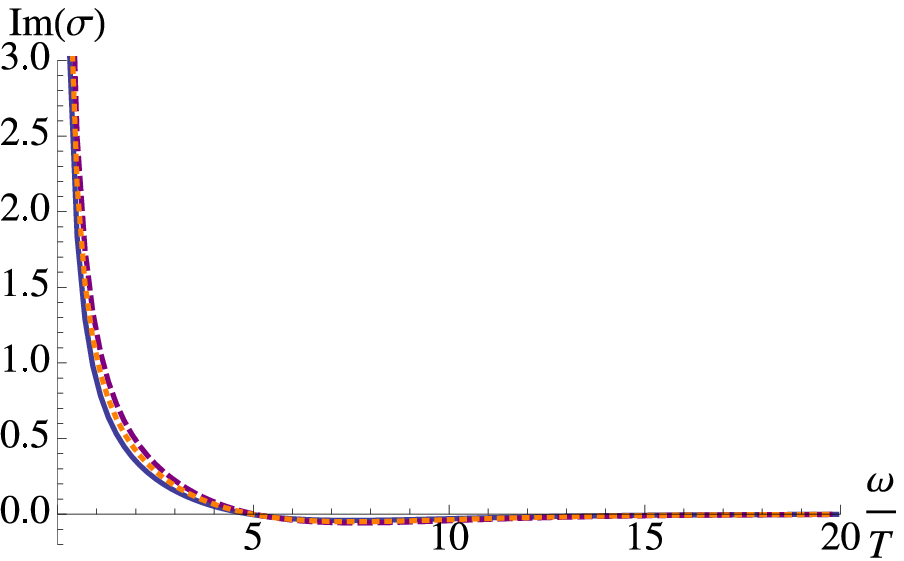}
\caption{The conductivity when $\Delta=2$ and $T/T_{\ast}=0.95$. The blue real lines are when $\tilde{c}=0.5$, the orange dotted lines are when $\tilde{c}=0$, and the purple dashed lines are when $\tilde{c}=-0.5$.}
\label{fig:hi_sigmaO2}
\end{figure}

It is expected from figure~\ref{fig:hi_condensate} that the effects of $J^t$ on the conductivity are significant only when $T/T_{\ast}$ is close to $1$. Hence, we would like to look into this region. We compute the conductivity when $T/T_{\ast}=0.95$, and the results are shown in figures~\ref{fig:hi_sigmaO1} and~\ref{fig:hi_sigmaO2}. We find that, in low frequencies, the real part of the conductivity increases if $\tilde{c}>0$. This agrees with the results of \cite{Hashimoto:2012pb}. However, we also find that the conductivity decreases if $\tilde{c}<0$. The imaginary part decreases if $\tilde{c}>0$, and increases if $\tilde{c}<0$. These shifts are due to the difference of $\phi$.

We thus find that it is simply the shift of $T_C$ that is relevant for the shifts of the conductivity in this model. This can be understood as follows: Owing to the shift of $T_C$, the magnitude of $\phi$ at some $T/T_{\ast}$ changes, and this results in the shift of the conductivity. The effects of this shift of $T_C$ disappear when $T/T_{\ast}$ is lowered, as differences in the condensate disappear. The mass gap in the conductivity appears in low temperatures. This is the same as the case that there is no impurity.

This result suggests that there is no dynamical effects of impurities in low temperature in this model. Indeed, the impurities in \eqref{hashimoto_action} is non-dynamical. The effects on the conductivity are present only near $T_{\ast}$, and this is due to the shift of $T_C$.  On the other hand, the action \eqref{Proca_action} includes the dynamical massive vector field, and there are nontrivial results in low temperatures. It should be noted that $T_C$ does not change in our model.

\section{Conclusion and discussions}
\label{sec:Conclusion}

In this paper, we considered a model of an s-wave holographic superconductor where a massive vector field for representing impurities was introduced. The model had an coupling of the gauge field $A_\mu$ and the massive vector field $B_\mu$, and was analyzed in the probe limit. We found that the massive gauge field was excited in the superconducting phase. A schematic phase diagram expected from our examination was given in figure~\ref{fig:dopingdiagram}. We also computed the optical conductivity for electromagnetic perturbation. When the coupling was sufficiently large, the mass gap in the conductivity disappeared. A resonance peak was found in the real part of the conductivity $\gamma_B$.
We also considered the case of normal phase in the RN-AdS black hole background, and found there was no effect on the conductivity. We discussed the model of \cite{Hashimoto:2012pb}, where impurities were proposed to be introduced non-dynamically by adding a bulk source term. We have shown that, in that model, the critical temperature for the superconductor phase were shifted due to the presence of the bulk source.

There are many things to be considered in the future. Several questions to be asked can be as follows. Perhaps the most straightforward extension of our work  is to include backreactions of the matter fields to gravity backgrounds, as considered in \cite{Hartnoll:2008kx}. We may also be able to consider non-relativistic deformations like the Lifshitz geometry \cite{Kachru:2008yh,Goldstein:2009cv}, which would be interesting as application to real-world condensed matter physics. We may also apply magnetic field to our model, and see the fate of the Meissner effect. The interaction we considered, $c F_{\mu\nu} G^{\mu\nu}$, is added in a bottom up way.  Since this interaction is crucial for the results in this paper, it would be interesting to derive this interaction from a top-down approach in string theory. This interaction might appear  in non-linear actions of D-branes. (Some related discussions can be found in \cite{Erdmenger:2011hp}.) Other interesting situations are to allow spatially inhomogeneous configuration as recently studied. In particular, realizing lattice structure \cite{Horowitz:2012ky,Horowitz:2012gs} is an interesting approach in modeling real-world materials. Another interesting case is to realize Josephson junctions \cite{Horowitz:2011dz}. We hope to come back to report on these topics.

\acknowledgments
We would like to thank Koji Hashimoto and Deog-Ki Hong for instructive discussions.
We also thank to APCTP for hospitality during the focus program ``Holography at LHC'', and
to Yukawa Institute Computer Facility where most of the numerical computations in this work were done. 
The work of SJS was supported by Mid-career Researcher Program through NRF grant No.~2010-0008456
and also supported by  the  NRF  grant through the SRC program CQUeST with grant number 2005-0049409.



\providecommand{\href}[2]{#2}\begingroup\raggedright\endgroup

\end{document}